\documentstyle[preprint,tighten,aps]{revtex}
\begin{document}
\title{Relativistic Hartree-Bogoliubov description of  
ground-state properties of Ni and Sn isotopes}
\author{G.A. Lalazissis$^1$, D. Vretenar$^2$,
and P. Ring$^1$}
\address{$^{1}$ Physik-Department der Technischen Universit\"at M\"unchen, 
Garching, Germany}
\address{$^{2}$ Physics Department, Faculty of Science, University of
Zagreb, Croatia}
\date{\today}
\maketitle
\begin{abstract}
The Relativistic Hartree Bogoliubov (RHB) theory is applied in
the description of ground-state properties of Ni and Sn isotopes.
The NL3 parameter set is used for the effective 
mean-field Lagrangian, and pairing correlations are described
by the pairing part of the finite range Gogny interaction D1S. 
Fully self-consistent
RHB solutions are calculated for the Ni ($28\leq N\leq 50$) 
and Sn ($50\leq N\leq 82$) isotopes. Binding energies, 
neutron separation energies, and proton and neutron $rms$ radii
are compared with experimental data. The model predicts 
a reduction of the spin-orbit potential with the 
increase of the number of neutrons. The resulting 
energy splittings between spin-orbit partners are
discussed, as well as pairing properties calculated 
with the finite range effective interaction in the $pp$ channel.
\end{abstract}
\vspace{1 cm}
%
\section {Introduction and outline of the model}

Relativistic models of nuclear
structure provide a
consistent framework in which the nuclear many-body system is
described in terms of interacting baryons and mesons.
Detailed properties of finite nuclei
along the $\beta$-stability line have been very
successfully described in the framework of relativistic
mean field models (for a recent review see Ref.~\cite{Rin.96}).  
In addition to the single-nucleon mean-field, pairing correlations
have to be taken into account
for a quantitative description of 
ground-state properties of
open-shell nuclei.
For nuclei close to the $\beta$-stability
line, pairing has been included in the relativistic
mean-field model in the form of a simple BCS
approximation~\cite{GRT.90}, with monopole pairing force
adjusted to the experimental odd-even mass differences.
Pairing correlations are 
also crucial for the description of deformations in heavy nuclei. 
However, as we move away from  the valley of 
$\beta$-stable nuclei, the 
ground-state properties calculated within the BCS 
approximation become unreliable. In particular 
properties that crucially depend on 
the spatial extensions of nucleon densities, as for 
example nuclear radii. The reason is that the BCS scheme
does not provide a correct description of the
scattering of nucleonic pairs from bound states to the
positive energy particle continuum~\cite{DFT.84,DNW.96},
and it leads to unbound systems. The solution, of course, 
is a unified description of mean-field and pairing
correlations, as for example in the framework of the
Hartree-Fock-Bogoliubov (HFB) theory. Within a 
non-relativistic approach to the nuclear many-body problem, 
HFB theory has a long and successful history of applications.
In particular, HFB theory in coordinate space~\cite{DFT.84}
has been used to describe properties of exotic nuclei 
with extreme isospin values, both on the neutron rich 
side~\cite{DNW.96}, and proton drip-line nuclei~\cite{Naz.96}.

The relativistic extension of the HFB theory was  
originally derived 
by Kucharek and Ring \cite{KR1.91}. Starting from 
the Lagrangian of quantum hadrodynamics, 
they have been able to show that 
the pairing correlations result from the one-meson exchange
($\sigma$-, $\omega$- and $\rho$-mesons). The  
Relativistic Hartree-Bogoliubov (RHB) model developed 
in Ref. \cite{KR1.91} is based on the 
Hartree approximation for the self-consistent mean field.
However, if for the one-meson exchange pairing interaction in RHB
one uses the coupling 
constants from standard parameter sets of the RMF model, 
the resulting pairing correlations are much too strong.
The repulsion produced by the 
exchange of vector mesons at short distances results in a 
pairing gap at the Fermi surface
that is by a factor three too large.
On the other hand, it has been argued in many applications of the 
Hartree-Fock-Bogoliubov theory that there is no real reason for
using the same effective forces in both the particle-hole and
particle-particle channel. In a first-order approximation,
the effective interaction contained 
in the mean-field $\hat\Gamma$ is
a $G$-matrix, the sum over all ladder diagrams. The
effective force in the $pp$ channel, i.e. in the pairing potential
$\hat\Delta$, should be the $K$ matrix, the
sum of all diagrams irreducible in $pp$-direction. However,
very little is known about this matrix in the relativistic
framework and only phenomenological effective forces have been 
used in the $pp$ channel of RHB. In the RHB calculations of 
Ref.~\cite{Gonz.96} pairing correlations have been described
by a two-body force of finite range of Gogny type \cite{BGG.84}, 
\begin{equation}
V^{pp}(1,2)~=~\sum_{i=1,2}
e^{-(( {\bf r}_1- {\bf r}_2)
/ {\mu_i} )^2}\,
(W_i~+~B_i P^\sigma 
-H_i P^\tau -
M_i P^\sigma P^\tau),
\end{equation}
with the parameters 
$\mu_i$, $W_i$, $B_i$, $H_i$ and $M_i$ $(i=1,2)$. For the D1S 
\cite{BGG.84} parameter set of the interaction the model was 
applied in the study of several isotope chains of 
spherical Pb, Sn and Zr nuclei. 
The pairing interaction is a sum of two Gaussians 
with finite range and properly chosen spin and isospin dependence. 
The Gogny force has been very carefully adjusted to the pairing 
properties of finite nuclei all over the periodic table. 
Its basic advantage 
is the finite range, which automatically guarantees a proper
cut-off in momentum space. In Refs.~\cite{PVL.97,LVP.98} we
have used RHB in coordinate space with the D1S Gogny interaction 
to study properties of light nuclear systems (C, N, O, F, Ne, Na, Mg) 
with large neutron excess. Self-consistent solutions were calculated
for the ground-states of a number of neutron-rich nuclei. Predictions
were obtained for the location of the neutron drip-line, reduction 
of the spin-orbit interaction, $rms$ radii, changes of surface 
properties, formation of neutron skin and neutron halo. The effective 
interactions in the $pp$ channel of the HFB theory were recently 
discussed in Ref.~\cite{DNW.96}. In particular, the role of finite 
range and the importance of density dependence were analyzed. 
Finite range forces have the obvious advantage of the automatic 
cut-off of high momentum components. On the other hand computations
with zero-range forces are much simpler, although an artificial 
energy cut-off has to be included. The density dependence of 
interactions in the pairing channel influences the spatial 
distributions of pairing densities and fields. By using 
interactions without density dependence,
strong pairing fields are produced in the volume of the 
nucleus. There is experimental evidence that pairing is a 
surface effect, and by including a density dependent component the 
pairing field can be made surface peaked. A fully self-consistent RHB model 
in coordinate space, with a density dependent interaction of 
zero-range (delta force),
has been used to describe the two-neutron halo in 
$^{11}$Li~\cite{MR.96}. 
\begin{equation}
V({\bf r_1}, {\bf r_2})= V_0 \delta({\bf r_1} - {\bf r_2})
{{1}\over {4}}\left (1 - \vec \sigma_1 \cdot \vec \sigma_2\right )
\left [ 1- {\rho(r)\over \rho_0} \right ]
\end{equation}

In going away from the valley of $\beta$-stable nuclei, the main 
problem of nuclear structure models becomes the extrapolation of effective
forces to nuclei with extreme isospin values. We have already applied
RHB with the Gogny force in the pairing channel to light neutron-rich
nuclei. However, many of these nuclei are still not accessible in 
experiments and therefore many of our results could not be compared 
with empirical data.  In order to make predictions for medium-heavy nuclei
at the neutron drip-line, we have to test available effective interactions
in detailed calculations of properties of neutron rich nuclei for 
which a comparison with experimental data is possible. In the present 
article we consider two sets of isotopes: Ni ($28\leq N \leq 50$)
and Sn ($50\leq N \leq 82$). Such an analysis will test the predictions
of effective forces in both the $ph$ and $pp$ channel over two 
major neutron shells. In addition to the effects of the pairing 
interaction, we are particularly interested in the behavior of the 
spin-orbit term of the effective potential as function of the 
neutron number. For light neutron rich nuclei, in
Ref.~\cite{LVR.97} we have shown that the magnitude of the spin-orbit 
potential is considerably reduced at the drip-line, resulting in 
much smaller energy splittings between spin-orbit partners. For the 
Ne isotopes this reduction was found to be around forty percent. 
Since it seems that at present only relativistic models include the correct
isospin dependence of the spin-orbit term in the mean-field 
potential, it would be important to study in medium-heavy 
nuclei the predicted spacings 
of single-neutron levels close to the Fermi surface. 

The details of the RHB model are given in 
Refs.~\cite{LVP.98,PVR2.97}. 
A short outline of essential features follows.
The ground state of a nucleus $\vert \Phi >$ is described
as vacuum with respect to independent quasi-particle operators,
which are defined by a unitary Bogoliubov transformation of the
single-nucleon creation and annihilation operators.  The
generalized single-nucleon Hamiltonian 
contains two average potentials: the self-consistent mean-field
$\hat\Gamma$ which encloses all the long range {\it ph}
correlations, and a pairing field $\hat\Delta$ which sums
up the {\it pp} correlations. The expectation value of the
nuclear Hamiltonian $< \Phi\vert \hat H \vert \Phi >$ is
a function of the hermitian density matrix
$\rho$, and the antisymmetric pairing tensor $\kappa$. The
variation of the energy functional with respect to $\rho$
and $\kappa$ produces the single quasi-particle 
Hartree-Fock-Bogoliubov equations~\cite{RS.80}.
The relativistic extension of the HFB theory was introduced
in Ref.~\cite{KR1.91}. In the Hartree approximation for
the self-consistent mean field, the Relativistic
Hartree-Bogoliubov (RHB) equations read
\begin{eqnarray}
\label{equ.2.2}
\left( \matrix{ \hat h_D -m- \lambda & \hat\Delta \cr
                -\hat\Delta^* & -\hat h_D + m +\lambda} \right) 
\left( \matrix{ U_k({\bf r}) \cr V_k({\bf r}) } \right) =
E_k\left( \matrix{ U_k({\bf r}) \cr V_k({\bf r}) } \right).
\end{eqnarray}
where $\hat h_D$ is the single-nucleon Dirac
Hamiltonian (\ref{statDirac}) and $m$ is the nucleon mass.
The chemical potential $\lambda$  has to be determined by
the particle number subsidiary condition, in order that the
expectation value of the particle number operator
in the ground state equals the number of nucleons. The column
vectors denote the quasi-particle wave functions, and $E_k$
are the quasi-particle energies. The Dirac Hamiltonian 
\begin{equation}
\label{statDirac}
\hat h_D~=~-i\mbox{\boldmath $\alpha$}
\cdot\mbox{\boldmath $\nabla$}
+\beta(m+g_\sigma \sigma)
+g_\omega \omega^0+g_\rho\tau_3\rho^0_3
+e\frac{(1-\tau_3)}{2} A^0
\end{equation}
contains the mean-field potentials of the isoscalar 
scalar $\sigma$-meson, the isoscalar vector $\omega$-meson
and the isovector vector $\rho$-meson. $A^0$ is the 
electrostatic potential. 
The RHB equations have to be solved self-consistently, with
potentials determined in the mean-field approximation from
solutions of Klein-Gordon equations
\begin{eqnarray}
\label{equ.2.3.a}
\bigl[-\Delta + m_{\sigma}^2\bigr]\,\sigma({\bf r})&=&
-g_{\sigma}\,\rho_s({\bf r})
-g_2\,\sigma^2({\bf r})-g_3\,\sigma^3({\bf r})   \\
\label{equ.2.3.b}
\bigl[-\Delta + m_{\omega}^2\bigr]\,\omega^0({\bf r})&=&
-g_{\omega}\,\rho_v({\bf r}) \\
\label{equ.2.3.c}
\bigl[-\Delta + m_{\rho}^2\bigr]\,\rho^0({\bf r})&=&
-g_{\rho}\,\rho_3({\bf r}) \\
\label{equ.2.3.d}
-\Delta \, A^0({\bf r})&=&e\,\rho_p({\bf r}).
\end{eqnarray}
for the sigma meson, omega meson, rho meson and photon
field, respectively. The spatial components 
\mbox{\boldmath $\omega,~\rho_3$} and ${\bf  A}$
vanish due to time reversal
symmetry. The equation for the sigma meson contains the 
non-linear $\sigma$ self-interaction terms~\cite{BB.77}.
Because of charge conservation, only the
3-component of the isovector rho meson contributes. The
source terms in equations (\ref{equ.2.3.a}) to
(\ref{equ.2.3.d}) are sums of bilinear products of baryon
amplitudes
\begin{eqnarray}
\label{equ.2.3.e}
\rho_s({\bf r})&=&\sum\limits_{E_k > 0} 
V_k^{\dagger}({\bf r})\gamma^0 V_k({\bf r}), \\
\label{equ.2.3.f}
\rho_v({\bf r})&=&\sum\limits_{E_k > 0} 
V_k^{\dagger}({\bf r}) V_k({\bf r}), \\
\label{equ.2.3.g}
\rho_3({\bf r})&=&\sum\limits_{E_k > 0} 
V_k^{\dagger}({\bf r})\tau_3 V_k({\bf r}), \\
\label{equ.2.3.h}
\rho_{\rm em}({\bf r})&=&\sum\limits_{E_k > 0} 
V_k^{\dagger}({\bf r}) {{1-\tau_3}\over 2} V_k({\bf r}).
\end{eqnarray}
where the sums run over all positive energy states. 
The pairing field
$\hat\Delta $ in (\ref{equ.2.2}) is defined
\begin{equation}
\label{equ.2.5}
\Delta_{ab} ({\bf r}, {\bf r}') = {1\over 2}\sum\limits_{c,d}
V_{abcd}({\bf r},{\bf r}') {\bf\kappa}_{cd}({\bf r},{\bf r}').
\end{equation}
where $a,b,c,d$ denote quantum numbers
that specify the single-nucleon states.
$V_{abcd}({\bf r},{\bf r}')$ are matrix elements of a
general two-body pairing interaction, and the pairing
tensor is defined 
\begin{equation}
{\bf\kappa}_{cd}({\bf r},{\bf r}') = 
\sum_{E_k>0} U_{ck}^*({\bf r})V_{dk}({\bf r}').
\end{equation}
The eigensolutions of Eq. (\ref{equ.2.2}) form a set of
orthogonal and normalized single quasi-particle states. The corresponding
eigenvalues are the single quasi-particle energies.
The self-consistent iteration procedure is performed
in the basis of quasi-particle states. The resulting quasi-particle
eigenspectrum is then transformed into the canonical basis of single-particle
states, in which the RHB ground-state takes the  
BCS form. The transformation determines the energies
and occupation probabilities of the canonical states.

For nuclear systems with spherical symmetry the fields
$\sigma(r),\,\omega^0(r),\,\rho^0(r),$ and $A^0(r)$ depend
only on the radial coordinate $r$.  The nucleon spinors
$U_k$ ($V_k$) in (\ref{equ.2.2}) are characterized by the
angular momentum $j$, its $z$-projection $m$, parity $\pi$
and the isospin $t_3=\pm {1\over 2}$ for neutron and
proton. The two Dirac spinors $U_k({\bf r})$ and $V_k({\bf r})$
are defined
\begin{eqnarray}
\label{spherspinor}
{U_k(V_k)}({\bf r},s,t_3)=
\pmatrix{ g_{U(V)}(r)\Omega_{j,l,m} (\theta,\varphi,s) \cr
        if_{U(V)}(r)\Omega_{j,\tilde l,m} (\theta,\varphi,s) \cr }
         \chi_\tau(t_{3}).
\end{eqnarray}
$g(r)$ and $f(r)$ are radial amplitudes, $\chi_\tau$ is the
isospin function, and 
$\Omega_{jlm}$ is the tensor product of the orbital and
spin functions
\begin{equation}
\Omega_{j,l,m} (\theta,\varphi,s)=\sum\limits_{m_s,m_l}
\bigl< {1\over 2}m_slm_l\big\vert jm\bigr> 
\chi_{{1\over 2} m_s} Y_{lm_l}(\theta,\varphi).
\end{equation}
The two-component functions
\begin{equation}
\Phi_U(r):=\pmatrix{g_U(r)\cr i f_U(r)}~~~~~~{\rm and}~~~~~~ 
\Phi_V(r):=\pmatrix{g_V(r)\cr i f_V(r)} ,
\end{equation}
are solutions of the Dirac-Hartree-Bogoliubov equations 
\begin{eqnarray}
\label{equ..2.22}
(\hat h_D(r)-m -\lambda )\Phi_U(r)+
\int_0^{\infty}dr'r'^2\Delta(r,r')\Phi_V(r') 
= E\Phi_U(r) \nonumber \\
(-\hat h_D(r)+m +\lambda )\Phi_V(r)+
\int_0^{\infty}dr'r'^2\Delta(r,r')\Phi_U(r') 
= E\Phi_V(r)
\end{eqnarray}
The self-consistent solution of the 
Dirac-Hartree-Bogoliubov integro-differential eigenvalue equations
and Klein-Gordon equations for the meson fields determines the 
nuclear ground state. In Refs.~\cite{PVL.97,LVP.98,LVR.97,PVR2.97}
we have used Finite Element Methods
in the coordinate space discretization of the coupled system
of equations. In order to correctly describe structure 
phenomena in exotic nuclei with extreme isospin values, as
for example regions of neutron halos with very diffuse neutron densities, 
the RHB equations have to be solved in coordinate space. However, 
for the spherical nuclear systems that we consider in the present
article this is not necessary \cite{DFT.84}.
As it was done in Ref.~\cite{Gonz.96}, 
the Dirac-Hartree-Bogoliubov equations and the equations for the 
meson fields are solved by expanding the nucleon spinors 
$U_k({\bf r})$ and $V_k({\bf r})$, 
and the meson fields in a basis of spherical harmonic oscillators 
for $N = 20$ oscillator shells~\cite{GRT.90}.
For odd-A isotopes we also include the  
blocking procedure \cite{RS.80} for neutron levels.  Strictly speaking, 
for these nuclei also the assumption about currents
is no longer valid, and in general one should include
the fields \mbox{\boldmath $\omega,~\rho_3$} and ${\bf  A}$.
However, the spatial components of the vector fields
do not significantly contribute to the ground state
properties investigated in the present work , and therefore 
they have not been included in our calculations. 
%
%
%
\section {Ground-states of Ni and Sn isotopes}
The details of ground-state properties of Ni ($28\leq N \leq 50$)
and Sn ($50\leq N \leq 82$) isotopes will depend on the choice 
of the effective mean-field Lagrangian in the $ph$ channel, as 
well as on the effective pairing interaction. Several
parameter sets  of the mean-field
Lagrangian have been derived that provide a satisfactory description 
of nuclear properties along the $\beta$-stability line. 
In particular,  NL1~\cite{RRM.86}, NL3~\cite{LKR.97}, and
NL-SH~\cite{SNR.93}.  The effective forces NL1 and NL-SH
have been used in many analyses to calculate properties of
nuclear matter and of finite nuclei, and have become
standard parameterizations for relativistic mean-field
calculations.  The effective interaction NL1 was also
used in the RHB+Gogny calculations of Ref. \cite{Gonz.96}.
More recently the parameter set NL3 has been derived 
\cite{LKR.97} by fitting ground state properties
of a large number of spherical nuclei.  Properties
calculated with the NL3 effective interaction are found to
be in very good agreement with experimental data for nuclei
at and away from the line of $\beta$-stability. 
In Ref. \cite{LR.98} it has been shown that constrained 
RMF calculations with the NL3 effective force reproduce the 
excitation energies of superdeformed minima relative to the 
ground-state in $^{194}$Hg and $^{194}$Pb. In the same work 
the NL3 interaction was also used for calculations 
of binding energies and deformation parameters of rare-earth
nuclei. We have used the NL3 parameter set in our study of light 
neutron-rich nuclei in Refs. \cite{PVL.97,LVP.98,LVR.97}.
For Ni and Sn the objective is to study 
how well does the NL3 effective force
describe systems with very different number of neutrons, 
without going to nuclei with extreme isospin values. Only 
if it were shown that the isospin dependence of NL3 is 
correct, this interaction could be used to make reliable 
predictions about medium-heavy drip-line nuclei. This could be 
especially important for Ni, since there is hope 
that drip-line isotopes might become accessible in 
experiments. Pairing being essentially a non-relativistic 
effect, we use the Gogny interaction in the $pp$ channel
with the parameter set D1S \cite{BGG.84}. Results obtained
with this effective force might also indicate the 
path one should take in deriving a fully relativistic 
theory of pairing, consistent with the modern mean-field 
Lagrangians.

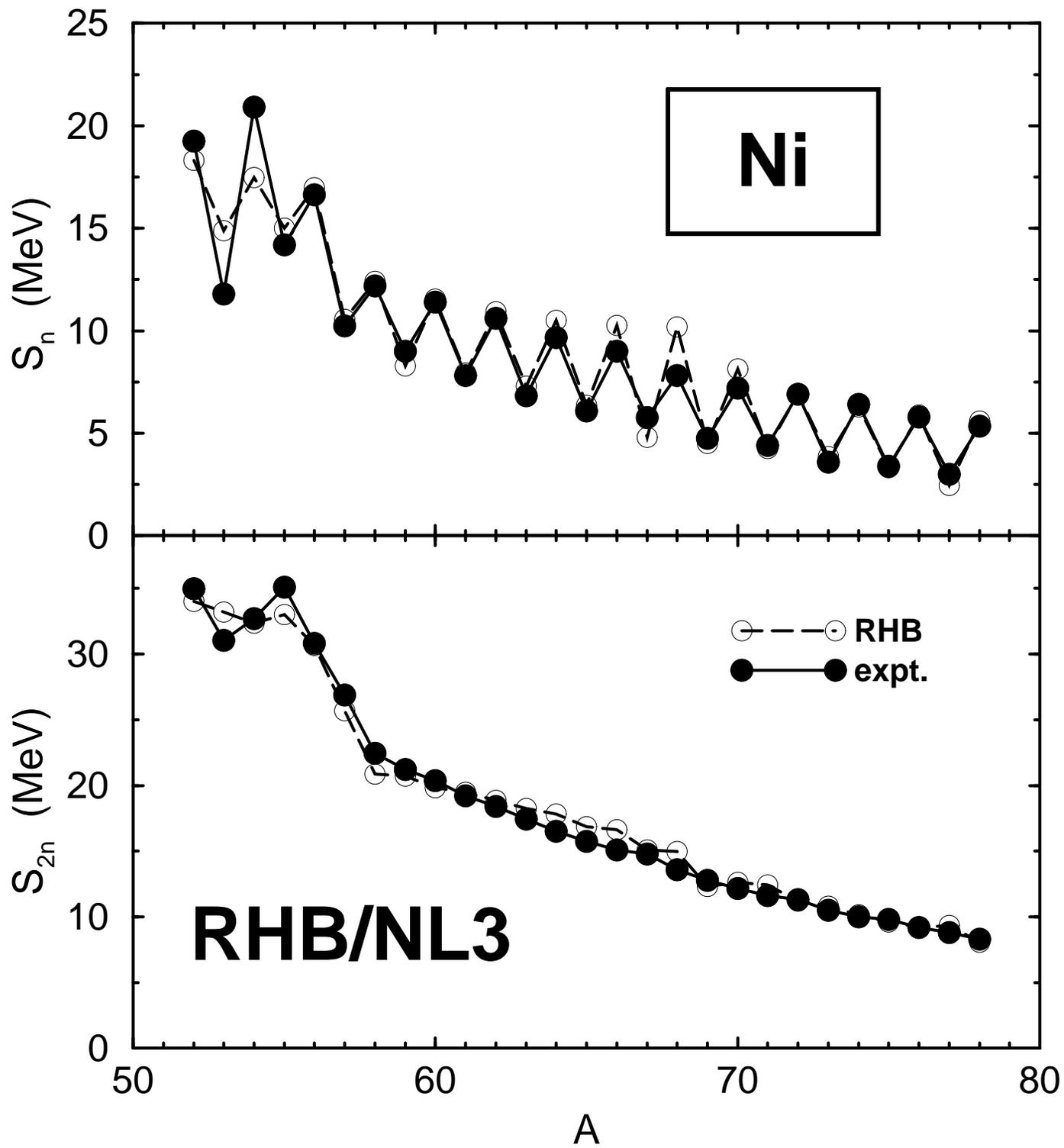
\begin{figure}
\caption{ One and two-neutron separation energies for Ni 
isotopes calculated in the RHB model and compared with 
experimental data.}
\label{figA}
\end{figure}

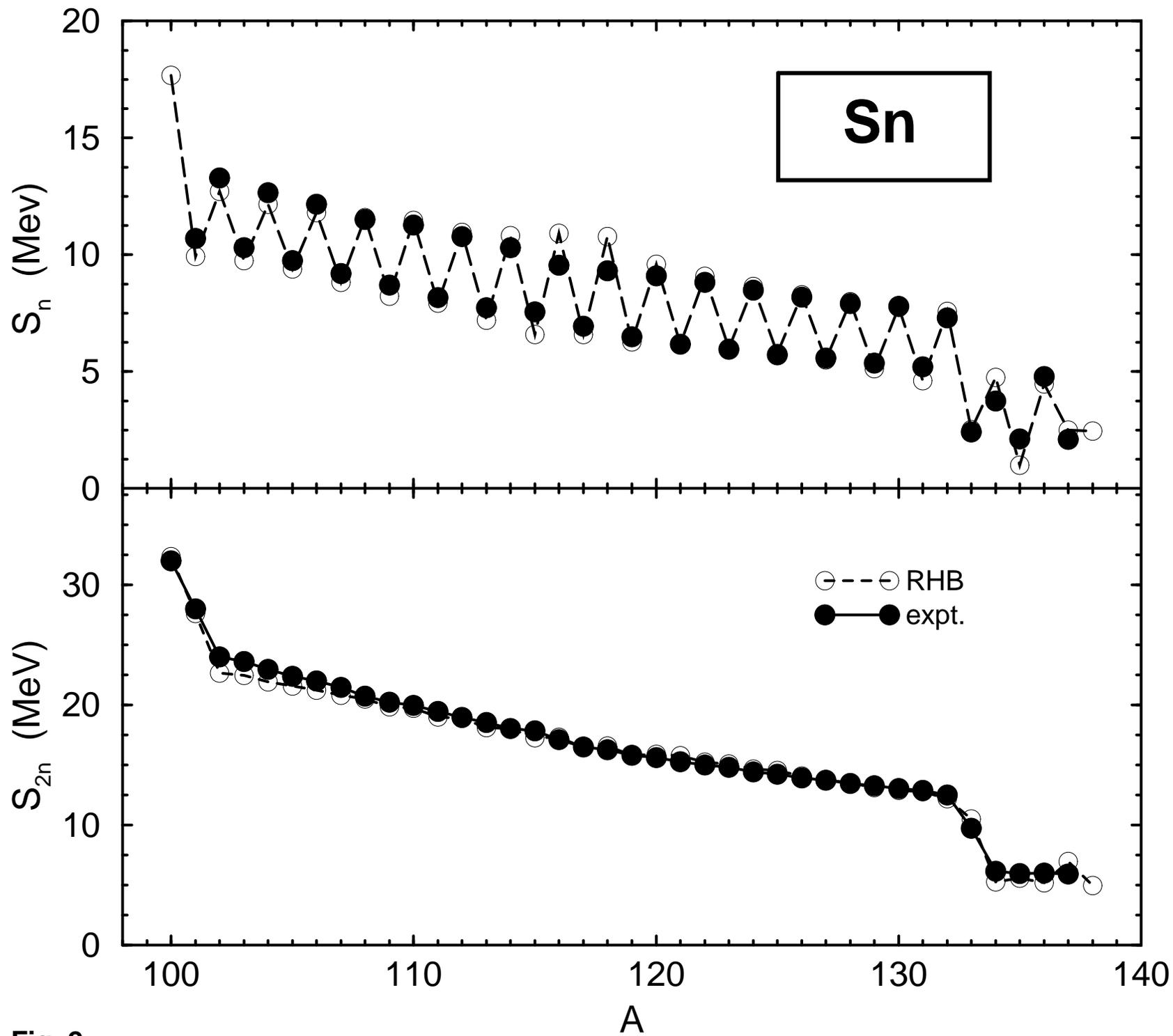
\begin{figure}
\caption{ Comparison between RHB model and experimental 
one and two-neutron separation energies for Sn isotopes.} 
\label{figB}
\end{figure}

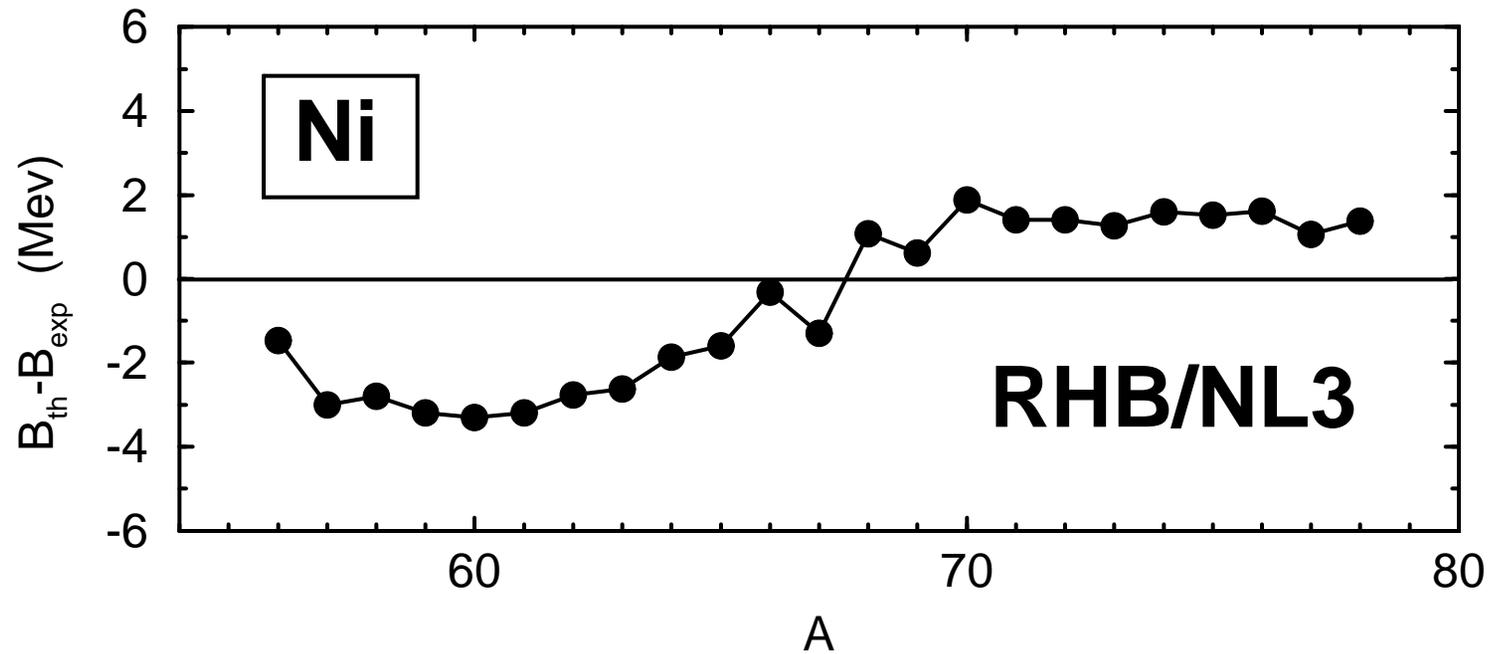
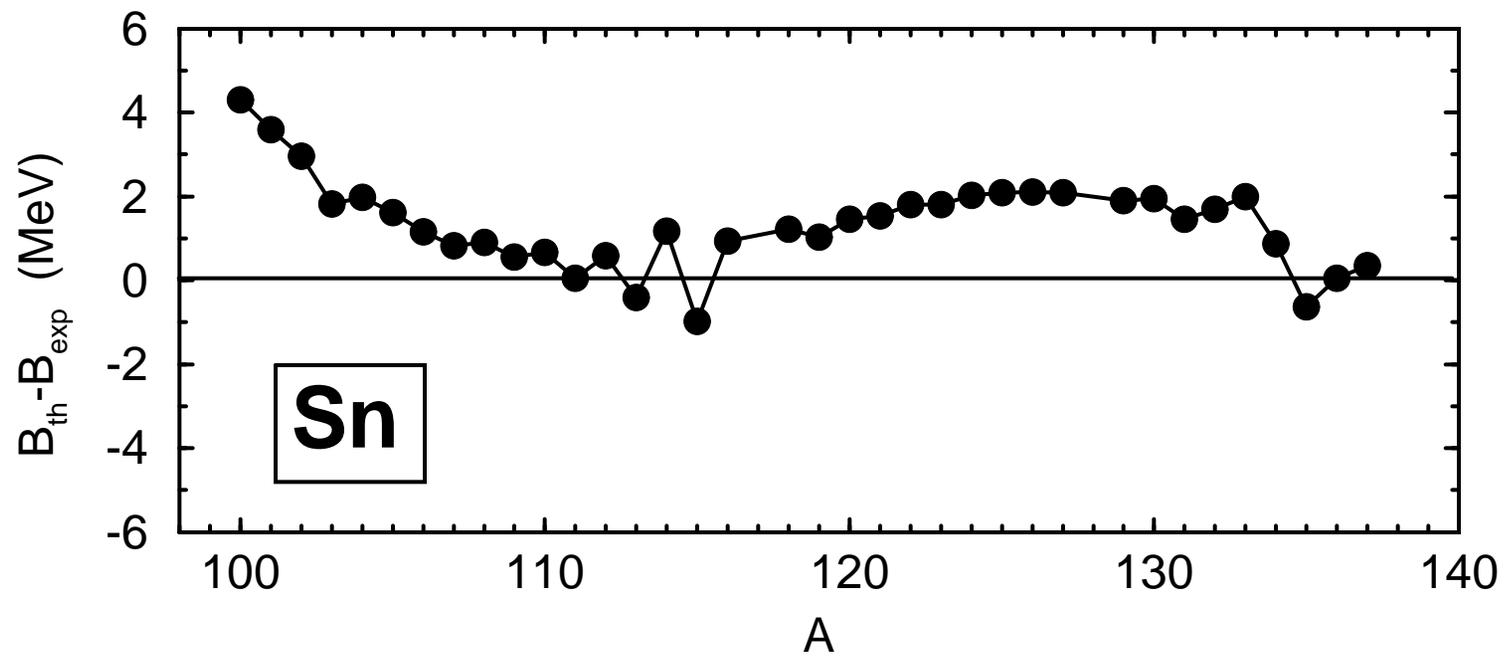
\begin{figure}
\caption{ Differences between RHB model and experimental 
binding energies for Ni and Sn isotopes.}
\label{figC}
\end{figure}

In Figs. \ref{figA} and \ref{figB} we display 
the one- and two-neutron separation 
energies
\begin{equation} 
S_n(Z,N) = B_n(Z,N) - B_n(Z,N-1)
\end{equation}
\begin{equation}
S_{2n}(Z,N) = B_n(Z,N) - B_n(Z,N-2)
\end{equation}
for Ni ($24\leq N \leq 50$) and Sn ($50\leq N \leq 88$) isotopes, respectively. 
The values that correspond to the self-consistent RHB ground-states are 
compared with experimental data and extrapolated values from 
Ref. \cite{AW.95}. The theoretical values reproduce in detail the 
experimental separation energies.  The model describes not only
the empirical values within one major neutron shell, but it also
reproduces the transitions between major shells. The results are
excellent for the region beyond  the shell closure at $N=82$ in Sn.  
The agreement with experimental data is somewhat worse for 
neutrons in the 1f$_{7/2}$ orbital in Ni isotopes, although the general 
trend is reproduced. However, these nuclei ($24\leq N \leq 28$) 
really belong to the proton-rich side and are not a primary objective 
of the present study. The total binding energies for Ni and Sn 
isotopes are compared with experimental values in Fig. \ref{figC}.
Except for the region around $^{60}$Ni and for $^{100-102}$Sn, 
the absolute differences between the calculated and experimental 
masses are less than 2 MeV. For Ni the model predicts weaker
binding for $N\leq 40$. Compared to experimental values,
the theoretical binding energies are 
$\approx$ 1 MeV larger for neutrons in the 1g$_{9/2}$ orbital
($40\leq N \leq 50$). For Sn 
isotopes the results of model calculation in general display an
overbinding. The differences are larger for $^{100-102}$Sn, 
but we expect that for these nuclei additional correlations should
be taken into account in order to get a better agreement with 
experimental data. In particular, proton-neutron pairing 
could influence the masses in this region. Proton-neutron 
short-range correlations are not included in our model.

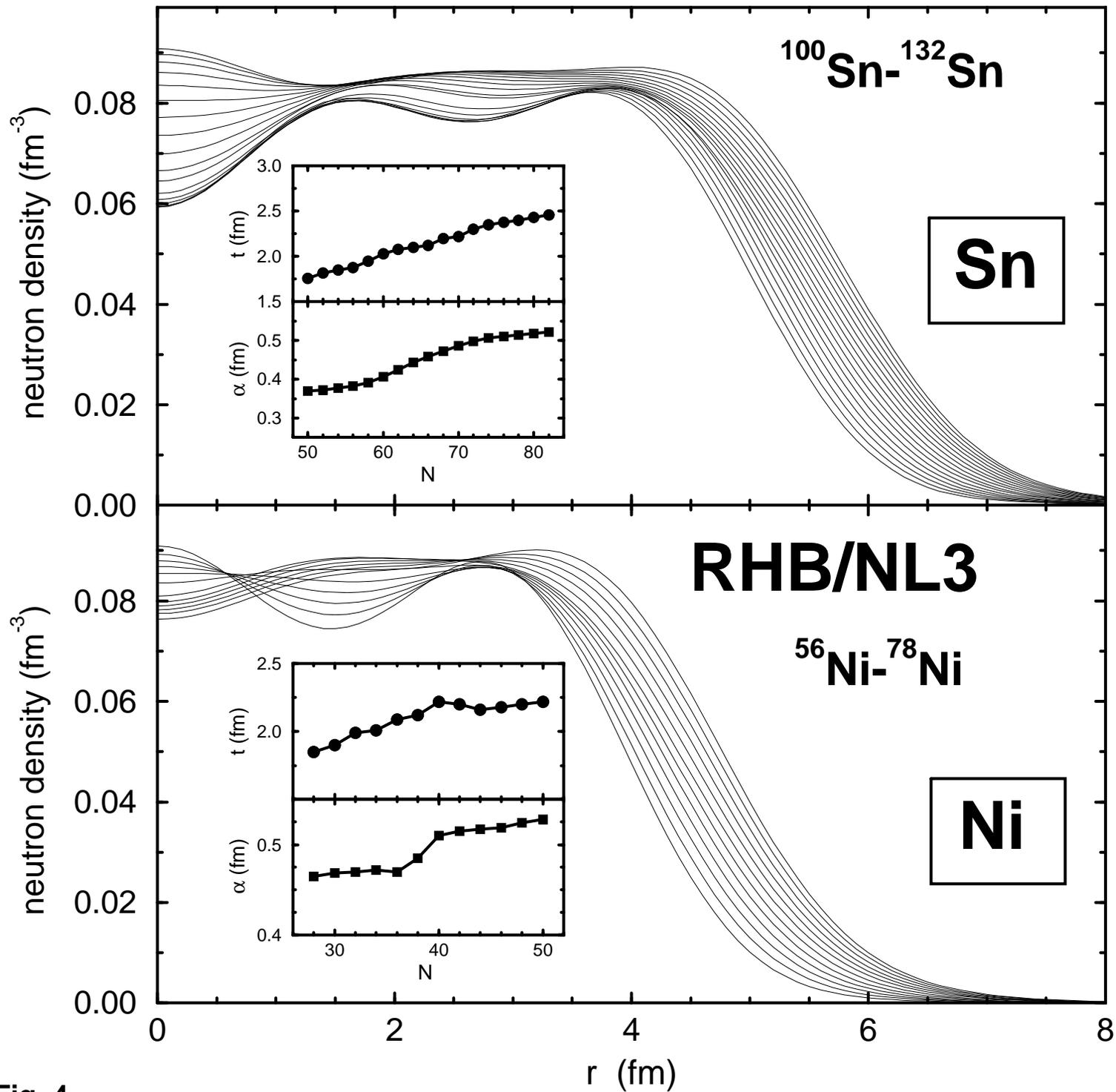
\begin{figure}
\caption{ Self-consistent RHB single-neutron density 
distributions for Sn ($50 \leq N\leq 82$)
and Ni ($28 \leq N\leq 50$) nuclei, calculated with the 
NL3 effective interaction.}
\label{figD}
\end{figure}

In Fig. \ref{figD} we show the self-consistent ground-state
neutron densities for the Sn and Ni nuclei. The 
density profiles display shell effects in the bulk and 
a gradual increase of neutron radii. In the
insert of Fig. \ref{figD} we include the corresponding values for
the surface thickness and diffuseness parameter. The
surface thickness $t$ is defined to be the change in radius
required to reduce $\rho (r) / \rho_0$ from 0.9 to 0.1
($\rho_0$ is the maximal value of the neutron density; because of
shell effects we could not use for  $\rho_0$ the density
in the center of the nucleus). The
diffuseness parameter $\alpha$ is determined by fitting the
neutron density profiles to the Fermi distribution
\begin{equation}
\rho (r) =  {\rho_0} \left (1 + exp({{r - R_0}\over 
\alpha})\right)^{-1} ,
\end{equation}
where $R_0$ is the half-density radius. By adding more 
units of isospin the value of the neutron surface thickness increases
and the surface becomes more diffuse. The increase in $t$ and 
$\alpha$ is more uniform in Sn, and both parameters increase 
approximately forty percent from $^{100}$Sn to $^{132}$Sn.
A somewhat smaller increase in the surface thickness is observed 
for Ni isotopes. The diffuseness parameter for Ni is essentially  
a step function: $\alpha \approx  0.45$ fm for $ N < 40$ and 
$\alpha \approx  0.50$ fm for neutrons in the 1g$_{9/2}$ orbital.
We will show that the observed changes in surface properties
result from the reduction of the spin-orbit term in the effective
single-nucleon potential. 

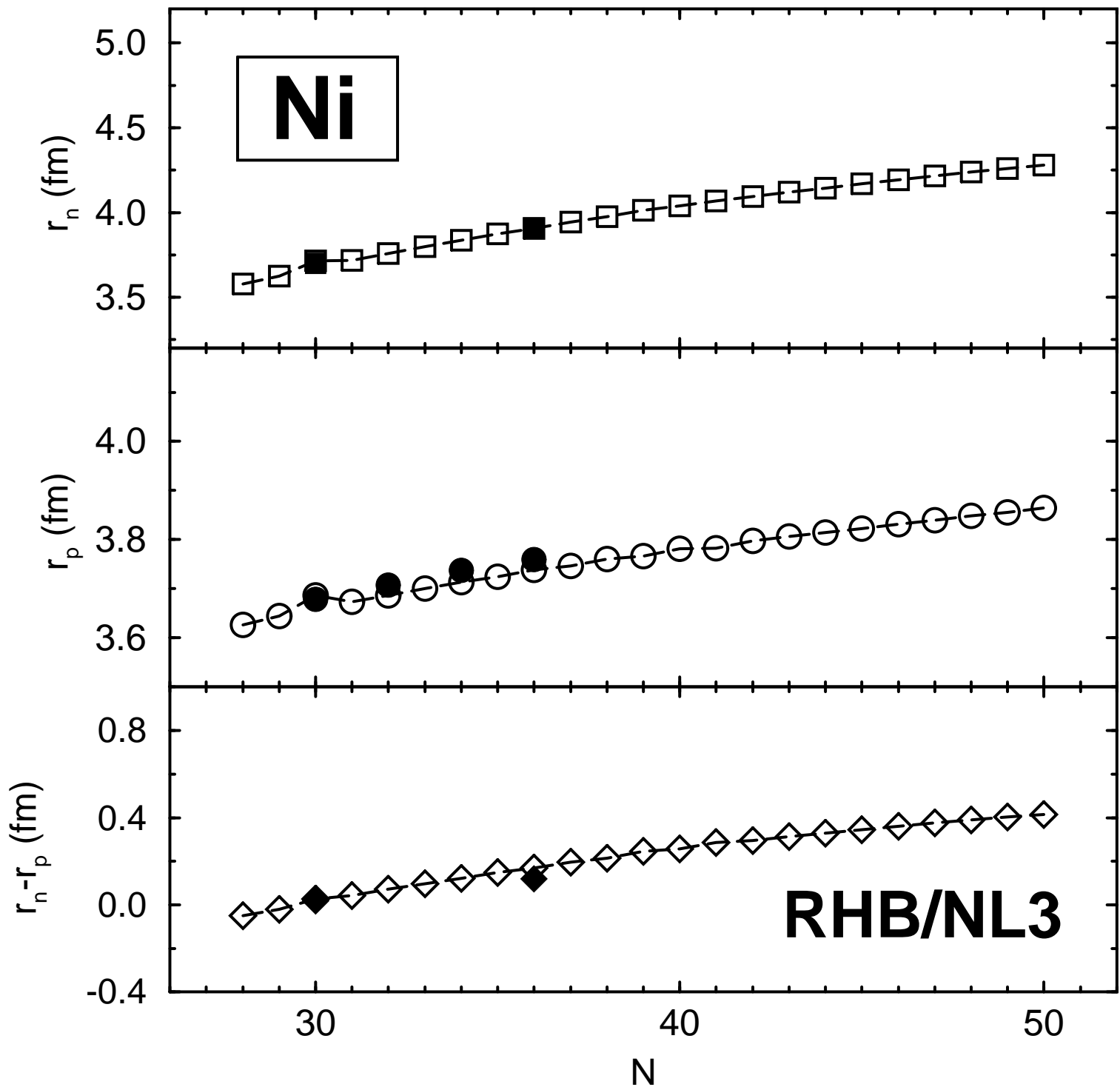
\begin{figure}
\caption{ Calculated neutron and proton $rms$ radii for Ni isotopes
compared with experimental data.}
\label{figE}
\end{figure}

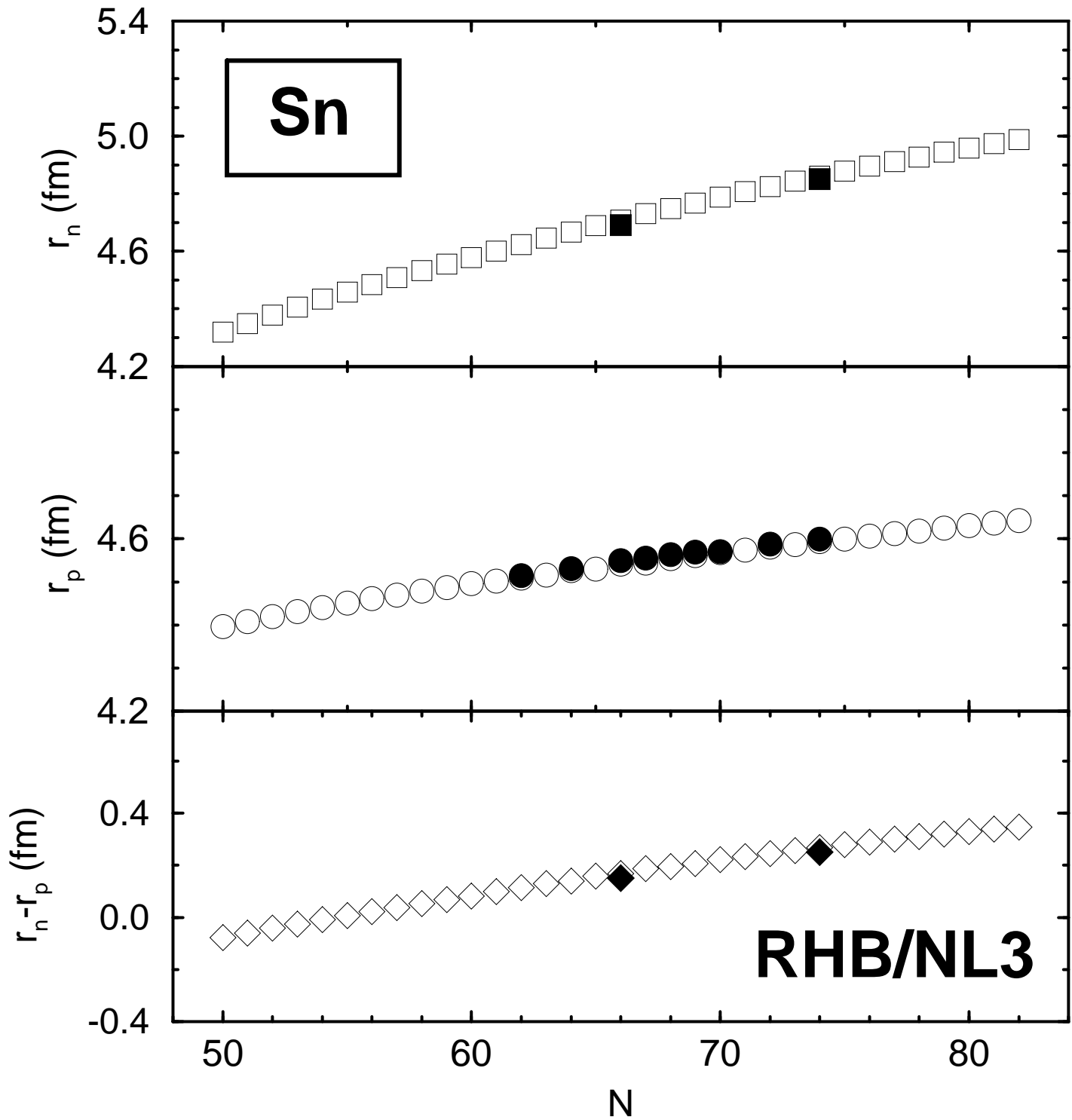
\begin{figure}
\caption{ Calculated and experimental neutron and proton 
$rms$ radii for Sn isotopes.}
\label{figF}
\end{figure}

In Figs. \ref{figE} and \ref{figF} we display the neutron and 
proton $rms$ radii for Ni and Sn isotopes, respectively. The 
calculated values are compared with experimental neutron 
radii from Ref. \cite{BFG.89}, and with data for proton radii 
from Ref. \cite{VJV.87}. In the lowest panels we also compare 
the differences $r_n - r_p$. We find an excellent agreement
between experimental data and values calculated with the 
NL3 effective force with the D1S Gogny interaction in the 
pairing channel. The model predicts a uniform increase 
of $rms$ radii with the number of neutrons. The neutron 
skin $r_n - r_p$ increases to approximately 0.4 fm at the 
closed shells $N=50$ for Ni, and $N=82$ for Sn. 

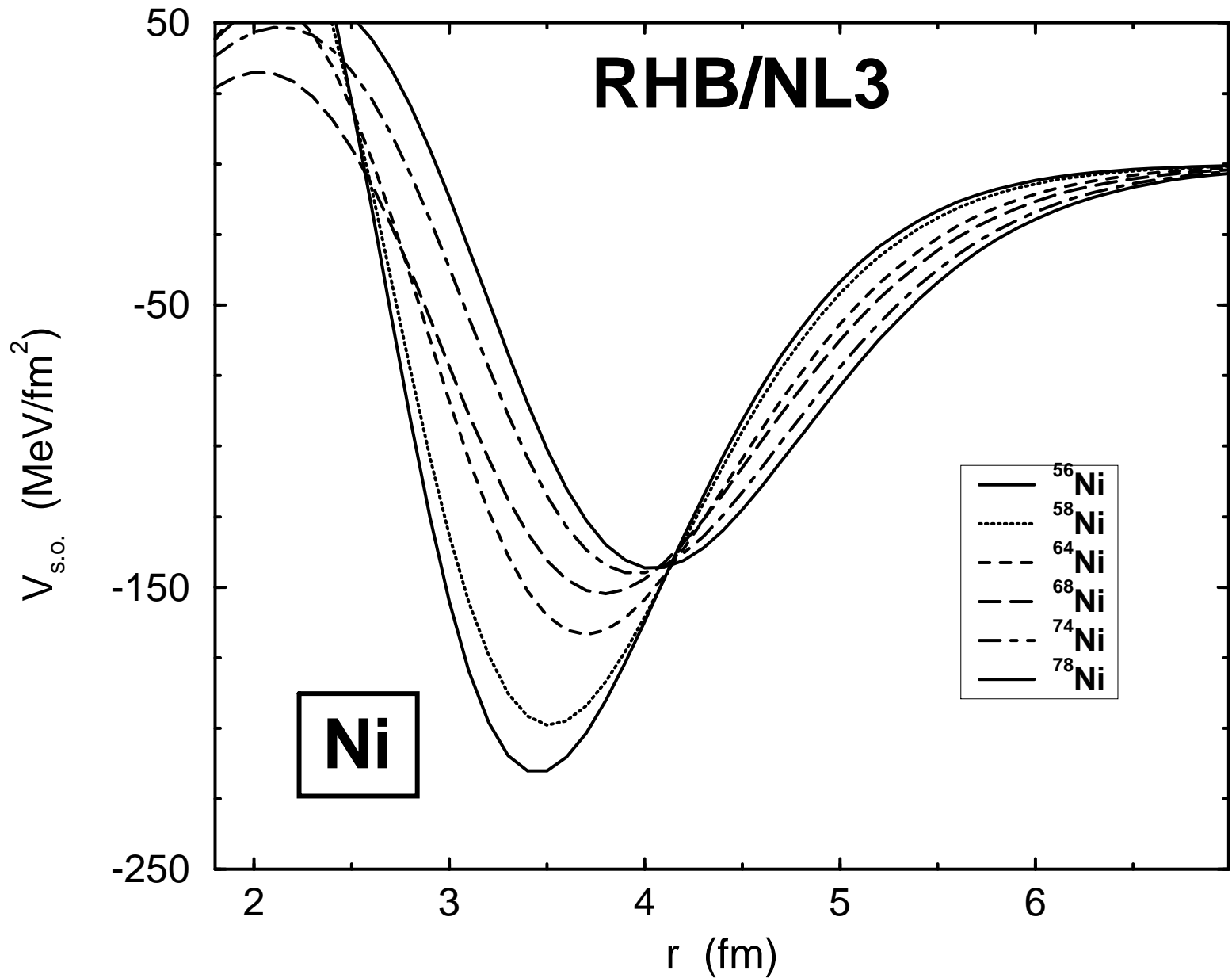
\begin{figure}
\caption{ Radial dependence of the spin-orbit term
of the potential in self-consistent solutions for the 
ground-states of Ni ($28 \leq N\leq 50$) nuclei.}
\label{figG}
\end{figure}

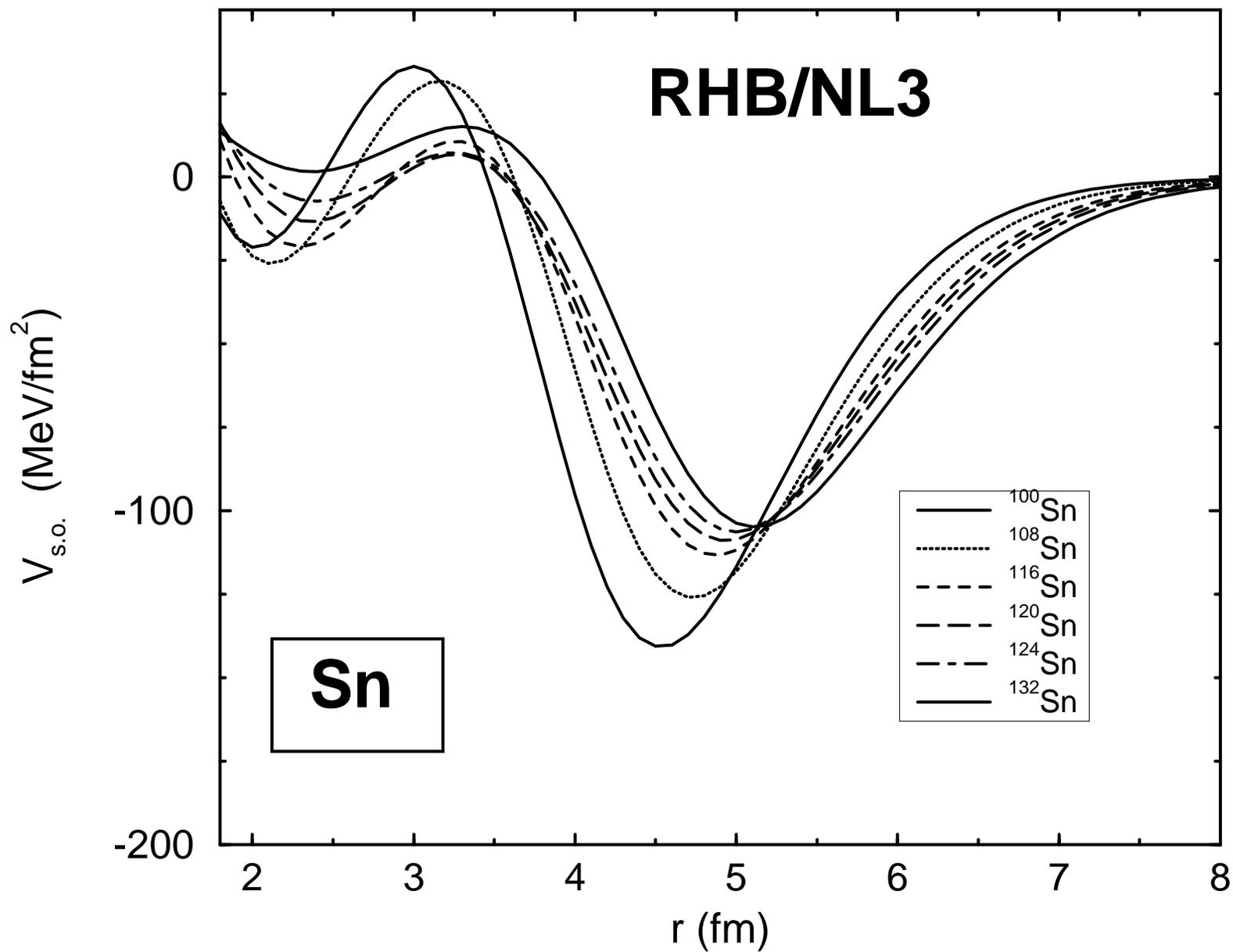
\begin{figure}
\caption{ Radial dependence of the spin-orbit term
of the potential in self-consistent solutions for the 
ground-states of Sn ($50 \leq N\leq 82$) nuclei.}
\label{figH}
\end{figure}

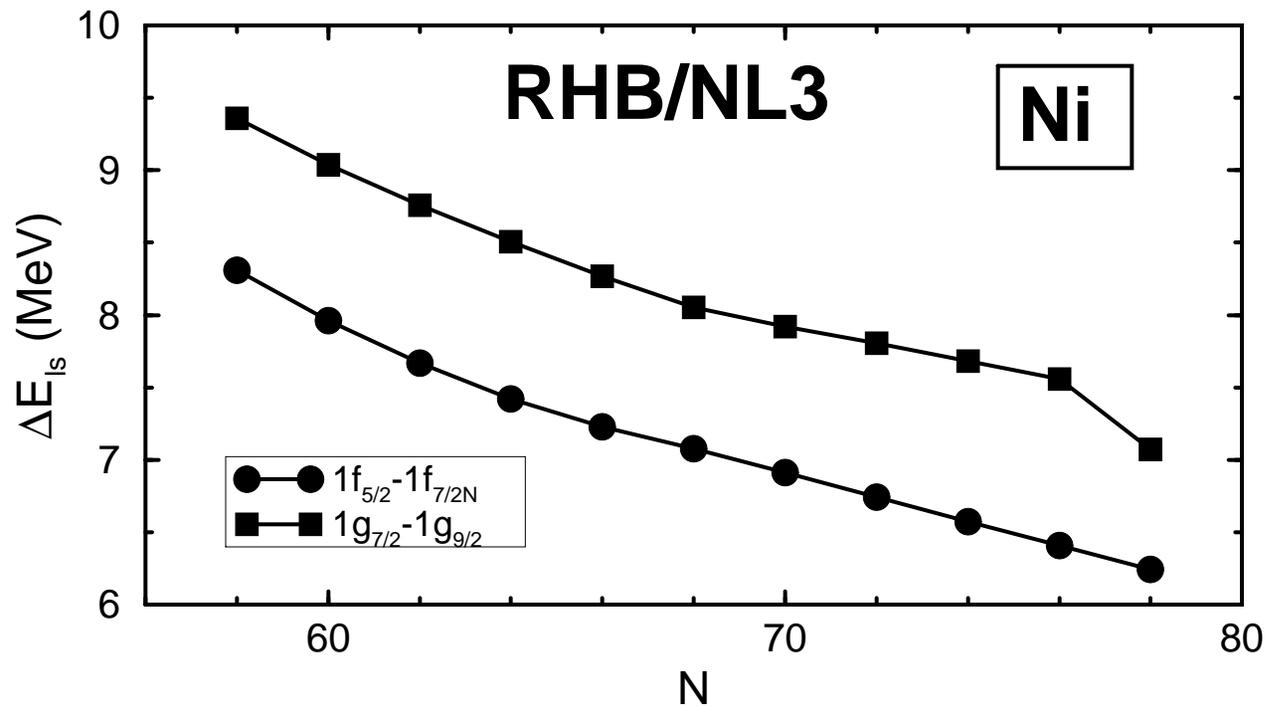
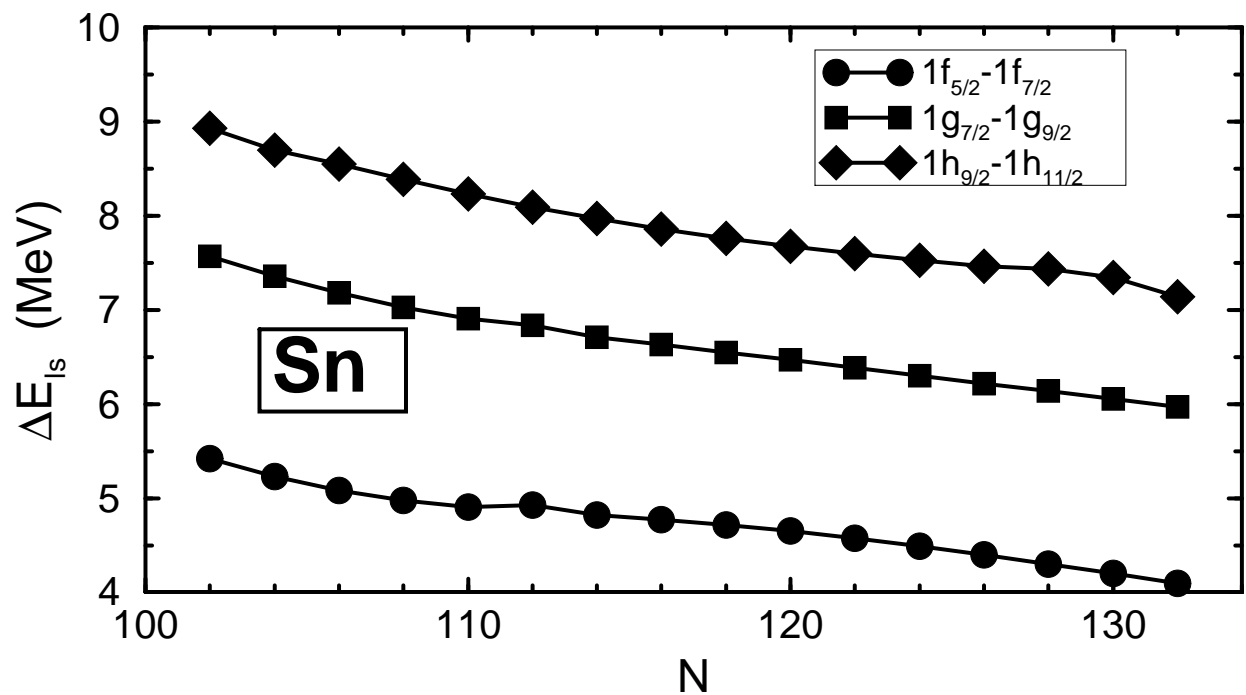
\begin{figure}
\caption{ Energy splittings between spin-orbit partners
for neutron levels in Ni and Sn isotopes, as functions
of neutron number.}
\label{figI}
\end{figure}

In Ref. \cite{LVR.97} we have shown that in the framework of the relativistic
mean-field model the magnitude of the spin-orbit term in the effective
single nucleon potential is greatly reduced for light neutron rich
nuclei. With the increase of the number of neutrons the effective 
spin-orbit interaction becomes weaker and this results in a reduction 
of the energy splittings for spin-orbit partners. The reduction in the 
surface region was found to be as large as $\approx 40 \%$ for Ne 
isotopes at the drip-line. The spin-orbit
potential originates from the addition of two large fields:
the field of the vector mesons (short range repulsion), and
the scalar field of the sigma meson (intermediate
attraction). In the first order approximation, and
assuming spherical symmetry, the spin orbit term can be
written as
\begin{equation}
\label{so1}
V_{s.o.} = {1 \over r} {\partial \over \partial r} V_{ls}(r), 
\end{equation} 
where $V_{ls}$ is the spin-orbit potential~\cite{KR2.91}
\begin{equation}
\label{so2}
V_{ls} = {m \over m_{eff}} (V-S).
\end{equation}
V and S denote the repulsive vector and the attractive
scalar potentials, respectively.  $m_{eff}$ is the
effective mass
\begin{equation}
\label{so3}
m_{eff} = m - {1 \over 2} (V-S).
\end{equation}
Using the vector and scalar potentials from the NL3
self-consistent ground-state solutions, we have computed
from~(\ref{so1}) - (\ref{so3}) the spin-orbit terms for the
Ni and Sn isotopes.  They are displayed in Figs. \ref{figG} 
and \ref{figH} as function of
the radial distance from the center of the nucleus. 
The magnitude 
of the spin-orbit term $V_{s.o.}$ decreases as we add more neutrons, i.e.
more units of isospin. If we compare $^{56}$Ni with $^{78}$Ni, 
in Fig. \ref{figG}, the reduction is $\approx 35\%$ in the surface region.
This implies a significant weakening of the spin-orbit interaction. 
The minimum of $V_{s.o.}$ is also shifted outwards, and this  
reflects the larger spatial extension of the scalar and vector densities,
which become very diffuse on the surface. The reduction of the 
spin-orbit term seems to be less pronounced in the Sn isotopes 
(Fig. \ref{figH}), and this indicates that the weakening of the 
spin-orbit interaction might be not that important in heavy nuclei.
The effect is reflected in the calculated  spin-orbit splittings 
of the neutron levels in the canonical basis
\begin{equation}
 \Delta E_{ls} = E_{n,l,j=l-1/2} - E_{n,l,j=l+1/2}, 
\end{equation}
In Fig. \ref{figI} we display the energy splittings of spin-orbit 
neutron partners for Ni and Sn, respectively. The calculated
spacings are shown  as function of the neutron number. We only
include the spin-orbit doublets for which one of the partners is 
an intruder orbital in a major shell.  These doublets display the 
largest energy splittings.  We notice in Fig. \ref{figI} that
the spacing between spin-orbit partners decreases with neutron number.
The effect is stronger in Ni than in Sn, and quantitatively the observed
decrease is consistent with the gradual weakening of the 
spin-orbit term shown in Figs. \ref{figG} and \ref{figH}. 

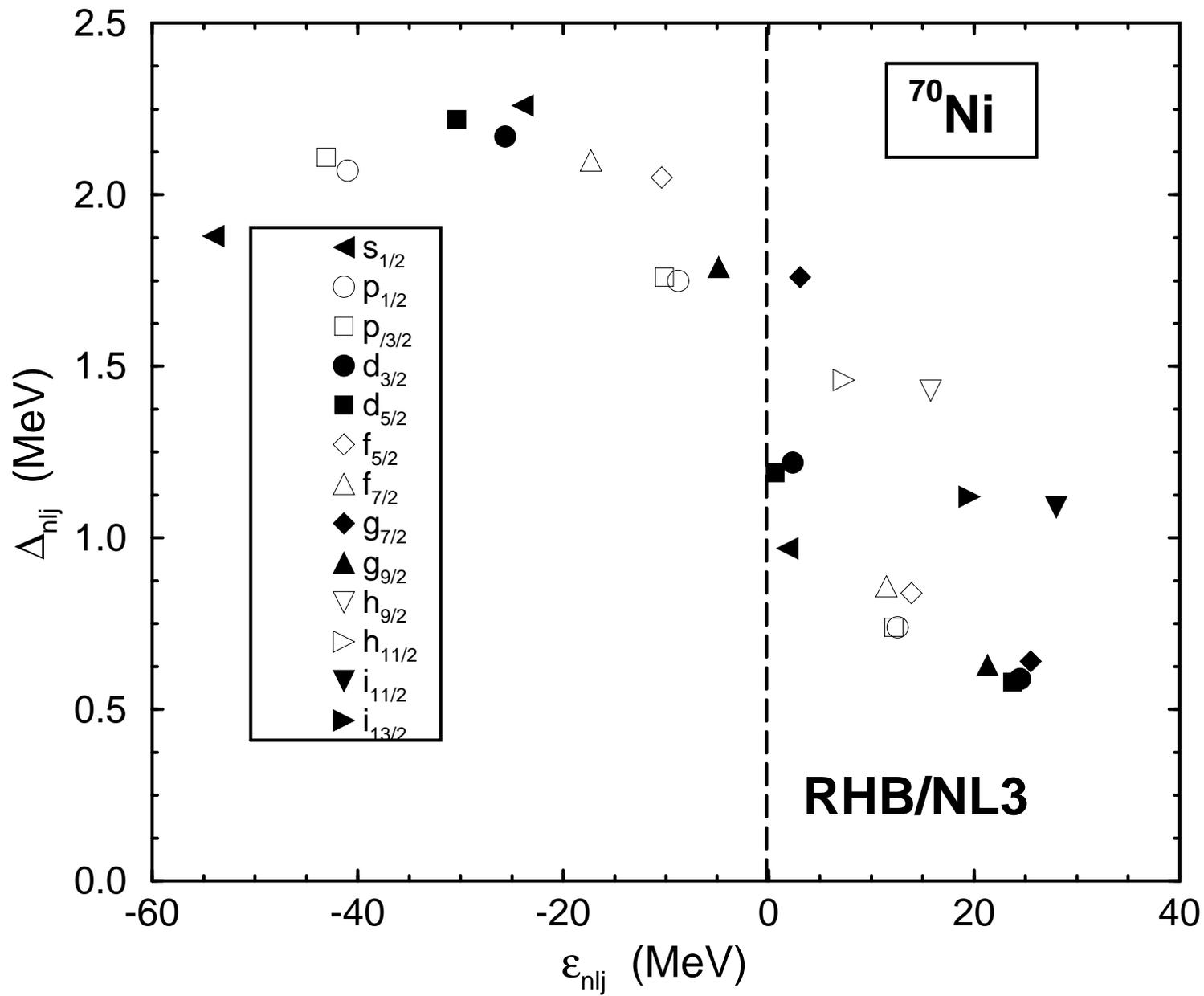
\begin{figure}
\caption{ Average values of the neutron canonical pairing gaps 
as functions of canonical single-particle energies for states that 
correspond to the self-consistent ground state of $^{70}$Ni. The NL3 
parametrization has been used for the mean-field
Lagrangian, and the parameter set D1S for the 
pairing interaction.}
\label{figJ}
\end{figure}

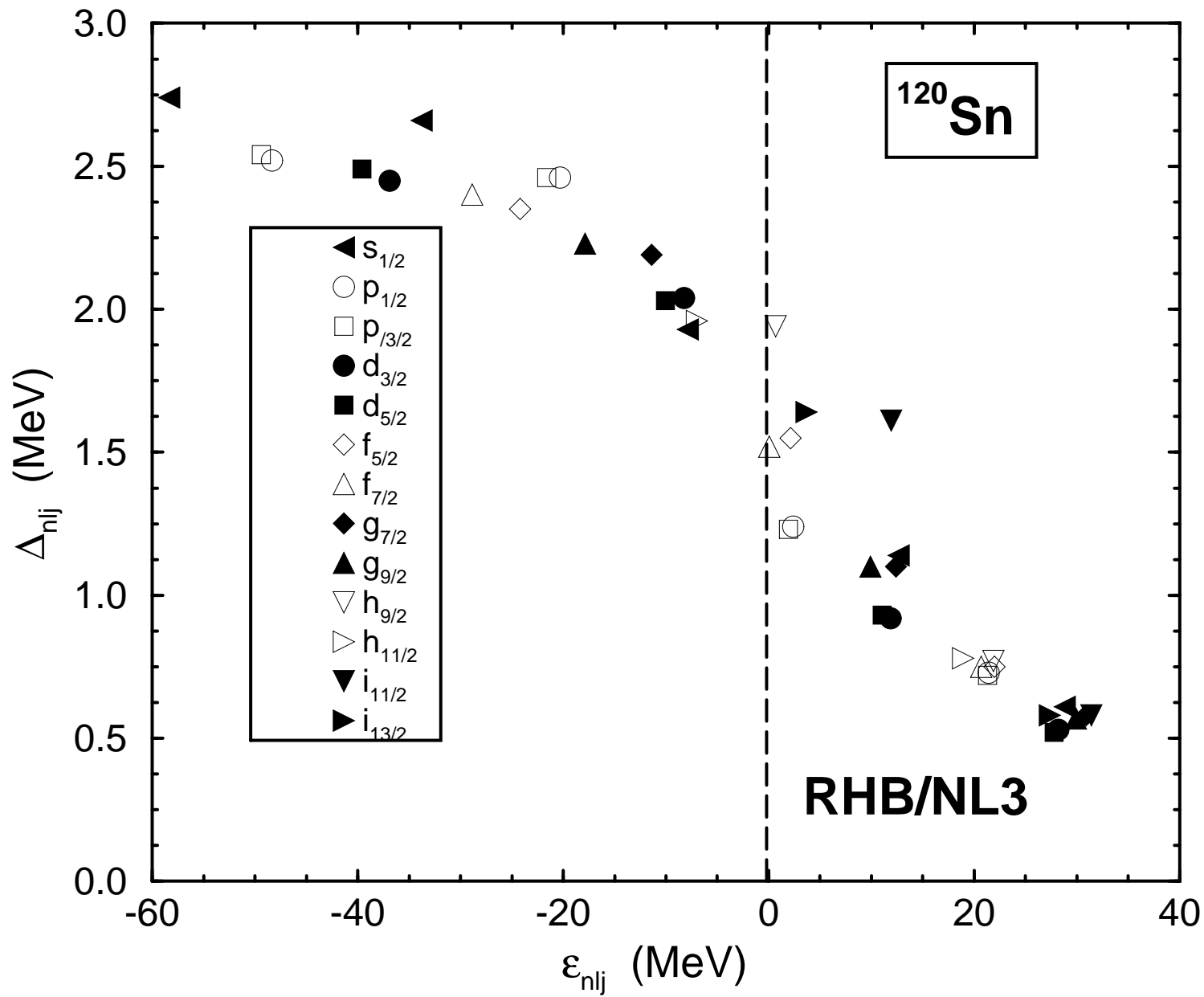
\begin{figure}
\caption{ Same as in the previous figure, but for the canonical states
that correspond to the ground state of $^{120}$Sn.}
\label{figK}
\end{figure}

In order to illustrate the properties of the interaction in the $pp$ channel, in 
Figs. \ref{figJ} and \ref{figK} we plot the average values 
of the neutron canonical pairing gaps $\Delta_{nlj}$ 
as functions of canonical single-particle energies. The gaps 
are displayed for canonical states that 
correspond to the self-consistent ground states of $^{70}$Ni  and
$^{120}$Sn, respectively. $\Delta_{nlj}$ are the diagonal matrix elements
of the pairing part of the RHB single-nucleon Hamiltonian in the 
canonical basis. Although not completely equivalent, $\Delta_{nlj}$
corresponds to the pairing gap in BCS theory. A very detailed 
discussion of HFB equations in the canonical basis can be found in 
Ref. \cite{DNW.96}. For $^{120}$Sn (Fig. \ref{figK}) we essentially 
reproduce the results calculated in Ref. \cite{DNW.96} with the Gogny 
HFB+D1S model. 
The calculated pairing gaps are almost identical, although there is 
a difference in the canonical single-neutron energies due to different
effective forces used in the $ph$ channel. While we employ the 
NL3 effective meson-exchange interaction in the $ph$ channel, in the 
calculation of Ref. \cite{DNW.96} the Gogny effective force has been 
used both in the $ph$ and $pp$ channel.
The pairing gaps display a uniform decrease with single-neutron energies.
This is related to the volume character of the Gogny interaction in the 
pairing channel. The values of the $\Delta_{nlj}$ for the s$_{1/2}$ orbitals 
are slightly larger, and the average value at the Fermi surface is around 
1.5 MeV.  It should be noted that for the HFB+SkP calculations of 
Ref. \cite{DNW.96} the values of the pairing gaps are strongly 
peaked at the Fermi surface. In addition, we have calculated the average 
canonical pairing gaps for $^{70}$Ni (Fig. \ref{figJ}). As compared 
with $^{120}$Sn, the $\Delta_{nlj}$ display a very similar behavior 
for deep-hole states (although for 1s$_{1/2}$ the pairing gap is considerably
smaller), but they decrease more slowly for canonical states in the 
single-neutron continuum. At the Fermi surface the average values are
between 1 and 1.5 MeV, somewhat smaller than in $^{120}$Sn.

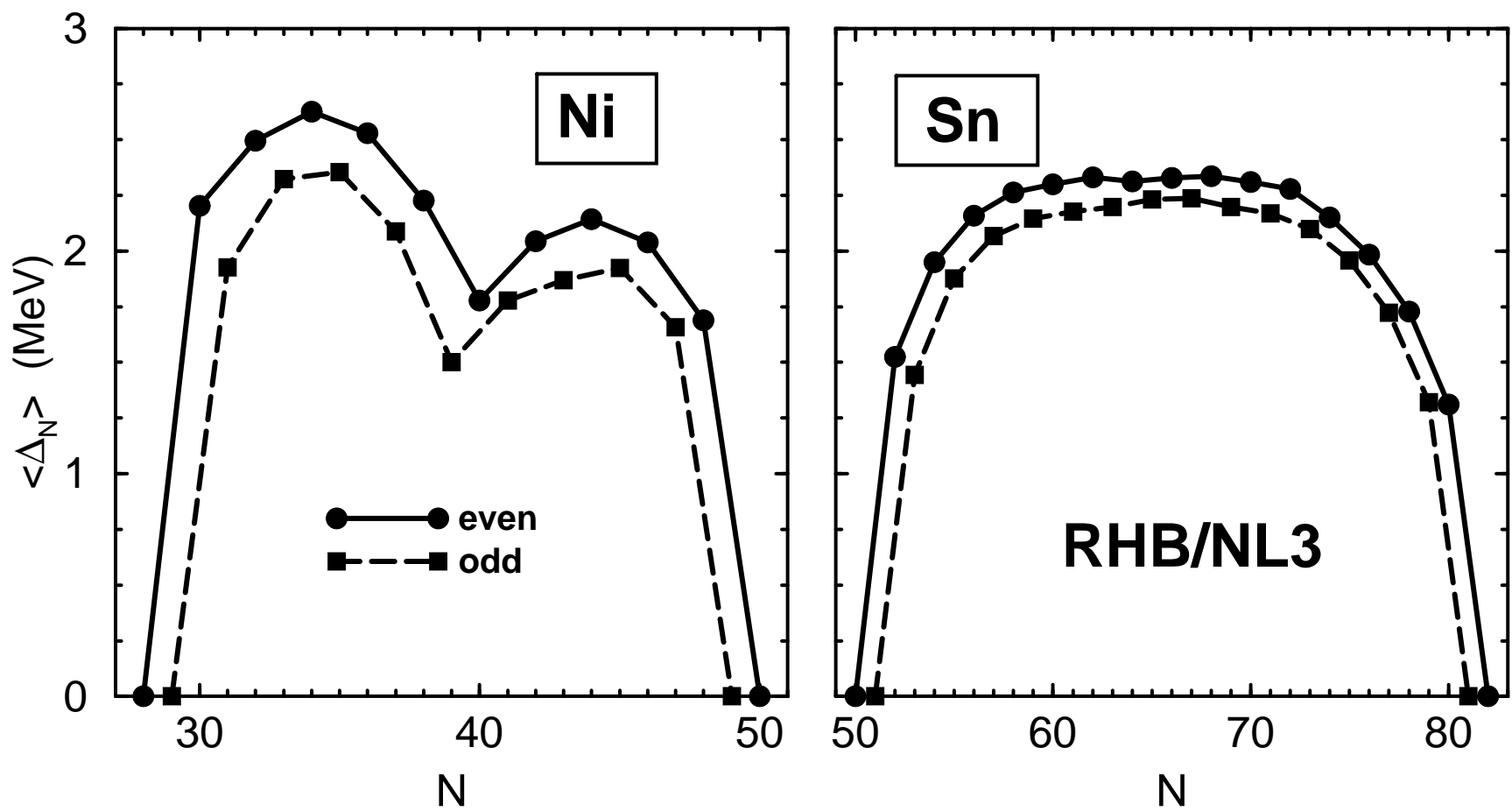
\begin{figure}
\caption{ Average neutron pairing gaps $< \Delta_N >$ 
for the Ni and Sn isotopes, as functions
of neutron number.}
\label{figL}
\end{figure}

Finally in Fig. \ref{figL} we compare the averages of the neutron 
pairing gaps for occupied canonical states
\begin{equation}
< \Delta_N > = {{\sum_{nlj} \Delta_{nlj} v_{nlj}^2}\over
                  {\sum_{nlj}  v_{nlj}^2}},
\label{ang}
\end{equation}
where $v_{nlj}^2$ are the occupation probabilities. 
The quantities $< \Delta_N >$ are 
plotted as functions of the neutron number for Ni and Sn isotopes. 
Solid lines correspond to even, and dashed lines to odd-neutron isotopes. 
$< \Delta_N >$ provides an excellent quantitative measure
of pairing correlations. The quasi-parabolic functional 
dependence on the number of neutrons reflects the 
increase of pairing correlations toward the middle of a major shell. Both for 
Ni and Sn the values of  $< \Delta_N >$ are above 2 MeV  in
the middle of the corresponding shells. As one would expect, the 
values for odd-neutron isotopes are somewhat lower. While in 
Sn the average pairing gaps do not provide any indication of 
additional shell effects, in Ni they very clearly indicate the 
shell subclosure at $N=40$. It is also interesting to note that 
pairing correlations are stronger in the subshell $28 \leq N \leq 40$, 
than for neutrons in the 1g$_{9/2}$ orbital.

In conclusion, in the present work we have performed 
a detailed analysis of ground-state properties 
of Ni ($28\leq N\leq 50$) and Sn ($50\leq N\leq 82$) nuclei
in the framework of the Relativistic Hartree Bogoliubov theory.
The NL3 parameter set has been used for the effective 
mean-field Lagrangian in the $ph$ channel,
and pairing correlations have been described
by the finite range Gogny interaction D1S. 
In a comparison with available experimental 
data, we have shown that the NL3 + Gogny D1S effective
interaction provides an excellent description of binding energies, 
neutron separation energies, and proton and neutron $rms$ radii.
The results indicate that this choice of model parameters 
might also be valid for nuclei with more extreme isospin 
values, i.e. medium-heavy nuclei at the drip-lines.
We have also discussed the predicted
reduction of the effective spin-orbit potential with the 
increase of the number of neutrons, as well as the resulting 
energy splittings between spin-orbit partners and 
modifications of surface properties.
Pairing properties calculated 
with the finite range effective interaction in the $pp$ channel
have been carefully analyzed. These results are particularly 
important since one of the main objectives of RHB theory  
should be a fully relativistic description of pairing 
correlations, consistent with modern mean-field Lagrangians.


\end{document}